\documentclass[twocolumn]{jpsj2} 
\usepackage{graphicx}
\usepackage{dcolumn}
\usepackage{bm}
\def\beq{\begin{equation}}
\def\eeq{\end{equation}}
\def\bes{\begin{split}}
\def\ees{\end{split}}
\def\beqa{\begin{eqnarray}}
\def\eeqa{\end{eqnarray}}

\newcommand{\km}{k_{\rm FM}}

\newcommand{\klr}{k_{\rm L(R)}}
\def\lsim{\lower -0.3ex \hbox{$<$} \kern -0.75em \lower 0.7ex 
          \hbox{$\sim$}}
\def\gsim {\raise.35ex\hbox{$>$}\kern-0.75em\lower.5ex\hbox{$\sim$}}

\newcommand{\rme}{{\rm e}}

\begin{document}

\title{Sign reversal of ac Josephson current in a ferromagnetic Josephson junction}

\author{S. Hikino$^{1}$, M. Mori$^{1}$, S. Takahashi$^{1}$, S. Maekawa$^{1,2}$}
\inst{	
$^{1}$Institute for Materials Research, Tohoku University, Sendai 980-8577, Japan\\
$^{2}$CREST, Japan Science and Technology Agency, Tokyo 102-0075, Japan
}


\date{\today}

\abst{
The ac Josephson effect in a ferromagnetic Josephson junction, which is composed of two superconductors separated by a ferromagnetic metal (FM), 
is studied by a tunneling Hamiltonian and Green's function method. 
We obtain two types of superconducting phase dependent current, i.e., Josephson current and quasiparticle-pair-interference current (QPIC). 
These currents change their signs with thickness of the FM layer due to the 0-$\pi$ transition characteristic to the ferromagnetic Josephson junction. 
As a function of applied voltage, the Josephson critical current shows a logarithmic divergence called the Riedel peak at the gap voltage, 
while the QPIC shows a discontinuous jump. 
The Riedel peak reverses due to the 0-$\pi$~transition and disappears near the 0-$\pi$~transition point. 
The discontinuous jump in the QPIC also represents similar behaviors to the Riedel peak. 
These results are in contrast to the conventional ones. 
}

\kword{ac Josephson effect, Riedel peak, ferromagnetic Josephson junction, superconductor, ferromagnet, 0-state, pi-state}
\maketitle
\section{Introduction}
The Josephson effect is a macroscopic quantum phenomenon involving phase coherence between two superconductors (SC's), 
and is characterized by a zero voltage current through a thin insulating barrier (dc Josephson effect). 
On the other hand, when a finite voltage is applied across the junction, an alternating current flows according to time-dependence of phase-difference, 
$\theta$, in two superconductors (ac Josephson effect), 
which gives two types of current as, $I_{\rm J1}=I_{\rm c1}\sin\theta$ and $I_{\rm J2}=I_{\rm c2}\cos\theta$ \cite{josephson1}. 
The former is the same as the dc Josephson current except for time-dependence of $\theta$, 
while the latter is a phase-dependent dissipative current inherent to the ac case. 
As a function of applied voltage, $V$, the amplitudes, $I_{\rm c1}$ and $I_{\rm c2}$, show singularities originating from the gap, $\Delta$, 
in an $s$-wave superconductor. 
The $I_{\rm c1}$ has a logarithmic divergence called the Riedel peak at the gap voltage, $V = 2\Delta/e$, 
due to the square root singularities in the density of states \cite{riedel, werthamer, hamilton, harris, buckner_sis, morita},  
while the $I_{\rm c2}$ shows a discontinuity at $V = 2\Delta/e$ and is zero below $V < 2\Delta/e$ at zero temperature\cite{werthamer, harris}. 
The $I_{\rm c2}$ is experimentally observed near the superconducting transition temperature, $T_{\rm c}$ \cite{soerensen}, 
in various types of Josephson junctions \cite{likharev, averin_sns, hurd_sns, gonnelli_mgb2, qf_sun, jacobs}. 

Recently, a Josephson junction with a ferromagnetic metal (FM), 
which is called the ferromagnetic Josephson junction, has been actively studied 
both experimentally and theoretically \cite{golubov_sfs1, buzdin_sfs1, ryazanov_sfs1, kontos_sfs1, sellier_sfs1, sellier_sfs2, bauer_sfs1, frolov_sfs1, bell_sfs1, born_sfs1, weides_sfs1, 
oboznov_sfs1, shelukhin_sfs1, mori_sfs1}. 
One of the most interesting phenomena in the ferromagnetic Josephson junction is the oscillation of the $I_{\rm c1}$ 
as a function of the thickness of the ferromagnetic film. 
The mechanism of the oscillation is 
similar to that of Fulude-Ferrell-Larkin-Ovchinnikov (FFLO) state\cite{fulde, larkin}. 
Cooper pairs penetrating into the FM acquire a finite center of mass momentum proportional to the magnetic exchange splitting, $h_{ex},$ between 
up- and down-spin bands. 
As a result, the pair correlation in FM oscillates 
as a function of the thickness of FM.  
If the thickness of the FM is about a half of the period of the oscillation, 
the current-phase relation is shifted by $\pi $ from that of a conventional Josephson junction (0-junction) 
like as, 
$I_{\rm c1}<0$.
This is called a $\pi $-junction, which has potential applications as a quantum bit\cite{qbit_sfs1, qbit_sfs2}. 
In addition, the $\pi$-junction is used in some experiments to measure a nonsinusoidal current-phase relation. 
Using the ac Josephson effect, Sellier $et$ $al$. experimentally evidenced the second harmonic term 
given by $\sin2\theta$
in the SC/FM/SC junction\cite{sellier_sfs2}. 
However, most of studies on the $\pi$ junctions have been so far focused on the dc Josephson effect. 
Studies on the ac Josephson effect in ferromagnetic $\pi$ junctions 
will open a new pathway of basic physics and will contribute to realize the quantum bit including SC/FM/SC junctions. 

In this paper, we study the ac Josephson effect in a SC/X/SC junction, where X is either a NM or a FM. 
Using a tunneling Hamiltonian and Green's function method\cite{aslamazov, awaka, melin}, 
we obtain two types of phase dependent current, i.e. the ac Josephson current ($I_{\rm J1}=I_{\rm c1}\sin\theta$) and the QPIC ($I_{\rm J2}=I_{\rm c2}\cos\theta$).  
In a SC/NM/SC junction, $I_{\rm c1}$ and $I_{\rm c2}$ monotonously decreases with the thickness of NM. 
As a function of applied voltage, $I_{\rm c1}$ shows a logarithmic divergence (the Riedel peak) at the gap voltage, $V=2\Delta /e$, 
while $I_{\rm c2}$ discontinuously  jumps at the same voltage.   
On the other hand, in a SC/FM/SC junction, $I_{\rm c1}$ and $I_{\rm c2}$ exhibit the strong dependence on the thickness of FM 
and changes their signs by crossing the 0-$\pi $ transition point due to the magnetic exchange splitting between the up- and down-spin bands in FM. 
In particular, it is predicted that the Riedel peak in $I_{\rm c1}$ disappears at the 0-$\pi $ transition and 
the 0-$\pi$ transition occurs in $I_{\rm c2}$ like as $I_{\rm c1}$. 

The rest of this paper is organized as follows. 
In \S2, we introduce 
the 
model Hamiltonian including the tunneling Hamiltonian, and explain the formulation to calculate the ac current by the thermal Green's function method.  
In \S3, the ac Josephson current and the QPIC in the SC/NM/SC and SC/FM/SC junctions are shown as functions of thickness and applied voltage. We compare these two types of junctions and discuss similarities and differences.  
Summary and discussion are given in \S4.
Below, $\hbar=1$ and $k_{\rm B}=1$ are used in the equations. 

\section{Tunneling Hamiltonian approach for ac Josephson current}
\subsection{Model Hamiltonian}
We consider a junction composed of a ballistic FM with the thickness $d$ 
and $s$-wave SC electrodes as shown in Fig. \ref{sfs}.  
A dc bias voltage,  $V$, is applied across the junction. 

The SCs and the FM are connected by tunneling Hamiltonian. 
The total Hamiltonian is given by
\begin{align}
H &=
	H^{\rm L}_{\rm SC}+H^{\rm R}_{\rm SC}+H_{\rm FM}+H_{\rm T},
\label{total H} \\
H^{\rm L }_{\rm SC}
	&= 
		\sum\limits_\sigma {\int {d \textnormal{\mathversion{bold}$r$} 
		\psi _{{\rm L} ,\sigma }^\dag
		\left[ { - \frac{1}{{2m}}\bigtriangledown ^{2}  - \mu \frac{{e{V}}}{{\rm{2}}}} \right]
		\psi _{{\rm L} ,\sigma } } } \nonumber \\
		& + \Delta {\rm{e}}^{i\theta _{{\rm{L}}} } \int {d \textnormal{\mathversion{bold}$r$} 
		\psi _{{\rm L} , \uparrow }^\dag 
		\psi _{{\rm L} , \downarrow }^\dag  }  + h.c., 
\label{bcs h}\\
H^{\rm R }_{\rm SC}
	&=
		\left({\rm L}\rightarrow {\rm R}, - \mu \frac{eV}{2} \rightarrow +\mu \frac{eV}{2} \right),
\label{bcs h2} \\
H_{\rm FM}
	&= 
		\sum \limits_\sigma {\int {d \textnormal{\mathversion{bold}$r$}
		\psi _{{\rm{FM,}}\sigma }^\dag 
		\left[ { - \frac{1}{2m}\bigtriangledown ^{2} - \mu  - i\sigma h_{ex} } \right]\psi _{{\rm{FM,}}\sigma } } },
\label{fm h} \\
H_{\rm T}
	&= 
		\sum \limits_\sigma  {\int\limits_{\textnormal{\mathversion{bold}$r$} \in {\rm{L}},\textnormal{\mathversion{bold}$r$}' \in {\rm{FM}}} 
		{d \textnormal{\mathversion{bold}$r$}d \textnormal{\mathversion{bold}$r$}'
		{T}_{{\textnormal{\mathversion{bold}$r$},\textnormal{\mathversion{bold}$r$}'}}
		\psi _{{\rm{L,}}\sigma }^\dag 
		\psi _{{\rm{FM,}}\sigma } } }  \nonumber \\
		 &+ \sum\limits_\sigma  {\int\limits_{\textnormal{\mathversion{bold}$r$} \in {\rm{R}},\textnormal{\mathversion{bold}$r$}' 
		\in {\rm{FM}}} {d \textnormal{\mathversion{bold}$r$}d \textnormal{\mathversion{bold}$r$}'
		{T}_{ \textnormal{\mathversion{bold}$r$},\textnormal{\mathversion{bold}$r$}' } 
		\psi _{{\rm{R,}}\sigma }^\dag
		\psi _{{\rm{FM,}}\sigma } } } \nonumber \\
		&+ h.c., 
\label{t h}
\end{align}
where $\psi _{i, \sigma } \equiv \psi _{i, \sigma } \left({\bm r} \right)$ is the electron field operator with the position~$\bm{r}$ in the region $i$~(=L, R, or FM) and the spin $\sigma~(=\uparrow ,\downarrow)$.
We adopt the BCS mean field Hamiltonian $H^{\rm L(R)}_{\rm SC}$ with the $s$-wave gap $\Delta$ and the phase variable $\theta _{\rm L(R)}$ in the left (right) SC. 
Then, the phase difference is given by $\theta \equiv \theta _{\rm L} - \theta _{\rm R}$. 
The electron mass and the chemical potential are denoted by $m$ and $\mu$, respectively. 
Note that $\mp eV/2$ is added in the Hamiltonian (\ref{bcs h}), since the applied voltage, $V$, imposes a chemical potential difference, $V/2$, at each boundary between FM and SC. 
The Hamiltonian of the FM, $H_{\rm FM}$, which has no impurity scattering, has the exchange energy, $ h_{ex}$. 
$H_{\rm T}$ is the tunneling Hamiltonian,  
whose matrix element is denoted by  
$T_{{\bm r},{\bm r'}}$ and has a finite value at the SC/FM boundary as, 
$T_{{\bm r},{\bm r'}}=
{T_{0}}\delta \left(\textnormal{\mathversion{bold}$r$}-\textnormal{\mathversion{bold}$r$}'\right)
\delta \left(\textnormal{\mathversion{bold}$r$}-\textnormal{\mathversion{bold}$r$}_{\rm L (R)}\right)$.  
$\textnormal{\mathversion{bold}$r$}_{\rm L (R)}$ 
is the position of the interface between the left (right) SC and the FM. 

We calculate the expectation value of a current operator,
\begin{align} 
\hat J &= 
		- ie\sum\limits_\sigma {\int\limits_{{\bm r} \in {\rm L}, {\bm r'} \in {\rm FM}}} d{\bm r} d{\bm r'}
		T_{{\bm r}, {\bm r}'}
		{\rm{e}}^{ - ieVt/2}
		\psi _{{\rm{L,}}\sigma }^\dag \psi _{{\rm{FM,}}\sigma } + h.c, 
\label{current op}
\end{align}
%
where $x$ 
involves
both time, $t$, and $\textnormal{\mathversion{bold}$r$}$. 
The superconducting phase dependent current is given by,
\begin{eqnarray}
I_{\rm J}\equiv \left< \hat J \right> = I_{{\rm{c1}}} \sin \theta  + I_{{\rm{c2}}} \cos \theta, 
\label{ij}
\end{eqnarray}
where 
$\langle \hat J \rangle$ is the expectation value of the current operator. 
The first term of eq. (\ref{ij}) is the Josephson current and 
$I_{\rm c1}$ is the Josephson critical current. The second term is called the quasiparticle-pair-interference current (QPIC). 

\subsection{Josephson critical current and QPIC formula}
\begin{figure}[t]
\begin{center}
\includegraphics[width=7cm]{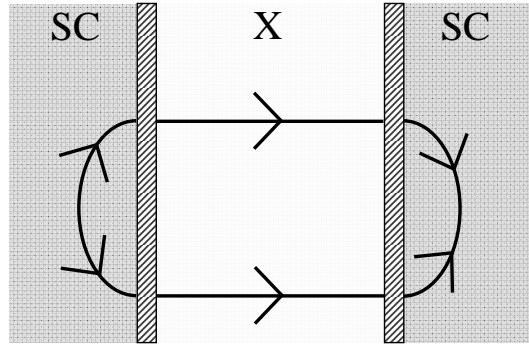}
\caption{Fourth order diagram with a tunneling matrix element contributing to $I_{\rm J}$. 
Insulating barrier is at interfaces between SC and X.}
\label{sfs}
\end{center}
\end{figure}
In the SC/FM/SC junction, the fourth order term of $\hat J $ with regard to 
${T}_{{\bm r}, {\bm r'}}$ is shown in Fig.$^{~} $\ref{sfs}. 
Detailed calculations are given in Appendices A and B. 
For $h_{\rm ex}/\mu $, $\omega _{n}/\mu <<1$, and temperature, $T$=0 [K], $I_{\rm c1}$ and $I_{\rm c2}$ are given by, 
\begin{align}
I_{\rm c1} &=
		\frac{\sigma _{0}\Delta_{0} ^{2}}{\pi e}
		\int_{\Delta_{0} }^\infty dE
		\frac{\Theta \left(\Delta_{0} -\left| E-eV\right| \right)}{\sqrt{E^{2}-\Delta_{0} ^{2}}\sqrt{\Delta_{0} ^{2}-\left(E-eV \right)^{2}}} \nonumber \\
		&\times \left\{ {\rm Ci}\left( \frac{2E-eV}{v_{\rm F}}d \right) \sin \left( \frac{2E-eV}{v_{\rm F}}d \right) \right. \nonumber \\
		&\left.-\cos \left(\frac{2E-eV}{v_{\rm F}}d \right) \left[{{\rm Si}\left(\frac{2E-eV}{v_{\rm F}}d \right)- \frac{\pi }{2}} \right] \right\} \nonumber \\
		&\times \cos \left(\frac{2h_{\rm ex}}{v_{\rm F}}d \right),
\label{ij1} \\
I_{\rm c2} &=
		\frac{\sigma_{0}\Delta_{0} ^{2}}{e}
		\int_{0}^{eV/2-\Delta_{0}} {dE} \nonumber \\
		&\times \frac{\Theta \left(eV-2\Delta_{0} \right)}
		{\sqrt{\left(E+eV/2 \right)^2-\Delta_{0} ^{2}}\sqrt{\left(E-eV/2 \right)^2-\Delta_{0} ^{2}}} \nonumber \\
		&\times \cos \left(\frac{2E}{v_{\rm F}}d \right) \cos \left(\frac{2h_{\rm ex}}{v_{\rm F} }d \right),
\label{ij2}
\end{align}
where $\Delta _{0}$ is the superconducting gap at $T$=0 [K], 
$\sigma _0  \equiv 16\pi e^{\rm{2}} {\rm{T}}_{\rm{0}}^{\rm{4}} N_{\rm{L}} (0)N_{\rm{R}} (0)\left[{ mv/(2\pi d)}
 \right]^{\rm{2}}$ is a constant determined by materials and interface, 
$N_{\rm {L(R)}}$ is the density of states in the left (right) lead at the Fermi level, $v_{\rm F}$ is the Fermi velocity, 
$v$ and $d$ are the volume and the thickness of the X, respectively.
Ci($x$) and Si($x$) are the cosine and sine integrals, respectively. 
When $h_{\rm ex}=0$, 
our formulation reproduces the current in the SC/NM/SC junction. 

For $d/\xi _{0}>>1$, eq. (\ref{ij1}) is simplified as 
\begin{align}
I_{\rm c1} &\simeq 
		\frac{\sigma _{0}\Delta_{0} ^{2}}{\pi e}
		\int_{\Delta_{0} }^{\infty }dE
		\frac{\Theta \left(\Delta_{0} -\left|E-eV \right| \right)}
		{\sqrt{E^{2}-\Delta_{0} ^{2}}\sqrt{\Delta_{0} ^{2}-\left(E-eV \right)^{2}}} \nonumber \\
		&\times \frac{v_{\rm F}}{2E-eV}
		\frac{1}{d},
\label{lm_ij1}
\end{align}
where $\xi _{0}\equiv v_{\rm F}/2\Delta_{0} $. 
It is found that $I_{c1}$ decays with $d$ in the power law as $1/d$. 
This behavior is consistent with the case of dc Josephson critical current in a double barrier Josephson junction\cite{mori_sfs1}. 
On the other hand, for $d\rightarrow 0$, eqs. (\ref{ij1}) and (\ref{ij2}) 
reproduce the current
in SC/I/SC junctions\cite{harris}. 
The integration in eqs. (\ref{ij1}) and (\ref{ij2}) is numerically carried out and results are shown in the next section.

\section{$d$- and $V$-dependence of $I_{\rm c1}$ and $I_{\rm c2}$}
In this chapter, we show thickness $d$ and bias voltage $V$ dependences of $I_{\rm c1}$ and $I_{\rm c2}$. 
First, the SC/NM/SC junction is discussed to see the difference between NM and FM cases clearly. 
Then, we will give detailed discussions on the SC/FM/SC junction. 
\subsection{{\rm SC/NM/SC} junction}
\begin{figure}[t]
\begin{center}
\includegraphics[width=6.5cm]{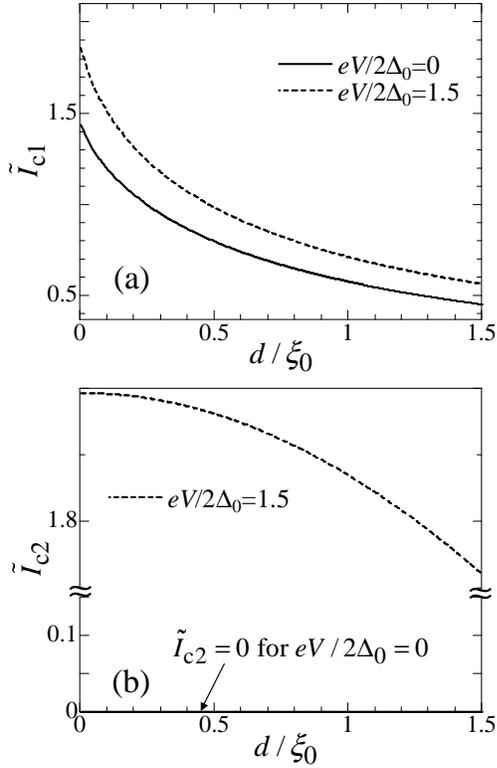}
\caption{(a) Normalized Josephson critical current, $\tilde I_{\rm c1}$, and (b) the quasiparticle-pair-interference current, $\tilde I_{\rm c2}$, 
as functions of $d/\xi _{0}$ for a SC/NM/SC junction ($h_{ex}=0$), where $\tilde I_{\rm c1}=I_{\rm c1}\pi e/\sigma_{0} \Delta_{0}$ and 
$\tilde I_{\rm c2}=I_{\rm c2}\pi e/\sigma_{0} \Delta_{0}$. 
}
\label{icd_sns}
\end{center}
\end{figure}

The thickness dependence of $I_{\rm c1}$ and $I_{\rm c2}$ is shown 
for several values of $V$ in Fig.~\ref{icd_sns}. 
In Fig.~\ref{icd_sns} (a), the vertical axis is the normalized $I_{\rm c1}$ and the horizontal axis is  
the thickness $d$ normalized by the coherence length $\xi _{0}$. 
It is found that $I_{\rm c1}$ shows a monotonic decrease as a function of $d$. 
For $d/\xi _{0}>>1$, $I_{\rm c1}$ decreases
in the power law as $1/d$ shown in eq. (\ref{lm_ij1}). 
In Fig.~\ref{icd_sns} (b), 
$I_{\rm c2}$ is plotted. 
It is found that $I_{\rm c2}$ decreases as a function of $d$ for $eV/2\Delta_{0} = 1.5$, while it is always equal to zero at $eV/2\Delta_{0} = 0$.
\newpage
\begin{figure}[t]
\begin{center}
\includegraphics[width=6.5cm]{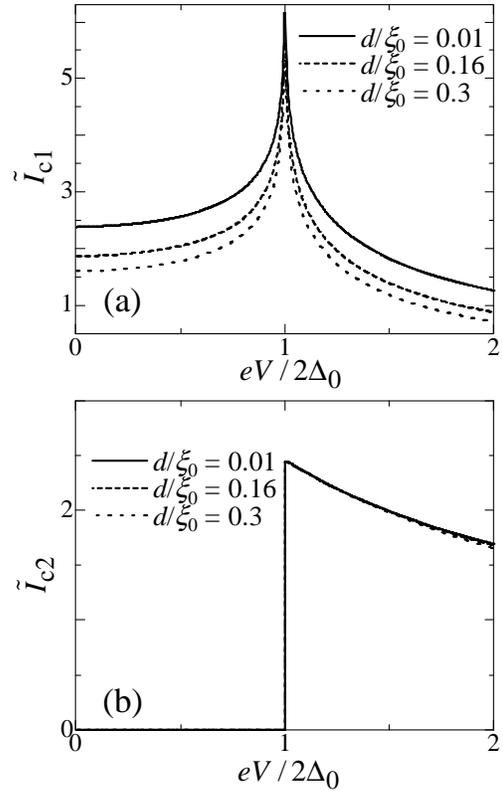}
\caption{(a) Normalized Josephson critical current $\tilde I_{\rm c1}$ and (b) 
quasiparticle-pair-interference current $\tilde I_{\rm c2}$ as functions of $eV/2\Delta _{0}$ in a SC/NM/SC junction
, where $\tilde I_{\rm c1}=I_{\rm c1}\pi e/\sigma_{0} \Delta_{0}$ and 
$\tilde I_{\rm c2}=I_{\rm c2}\pi e/\sigma_{0} \Delta_{0}$. 
}
\label{icv_sns}
\end{center}
\end{figure}
The $I_{\rm c1}$ in the SC/NM/SC junction is shown
for several values of $d$ 
in Fig.~\ref{icv_sns} (a). 
The vertical axis is the normalized $I_{\rm c1}$ and the horizontal axis is the normalized voltage, $eV/2\Delta _{0}$. 
In Fig.~\ref{icv_sns} (a), $I_{\rm c1}$ shows the Riedel peak at the gap voltage similar to that in a SC/I/SC junction. 
In this system, the Riedel peak exhibits 
weak
dependence on $d$. 
The $I_{\rm c2}$ is shown in Fig.${~}$\ref{icv_sns} (b). 
It is found that $I_{\rm c2}$ has discontinuity at $eV/2\Delta _{0}=1$. 
The behavior of $I_{\rm c2}$ is the same as the case of SC/I/SC and $I_{\rm c2}$ exhibits 
very weak
dependence on $d$. 

Here, we examine the above results in details.
$I_{\rm c1}$ shows the monotonic decrease as a function of $d$, 
which represents a decoherence of the Cooper pair penetrating into the NM. This behavior is 
equivalent 
to the case of dc Josephson effect. 
On the other hand, the behavior of $I_{\rm c2}$ is quite different from that of $I_{\rm c1}$,
since
$I_{\rm c2}$ has a finite value 
only
when $V$ is larger than the gap voltage as shown by Fig.~\ref{icv_sns} (b). 
Therefore, it is interpreted as quasiparticles in the band below the superconducting gap contribute to $I_{\rm c2}$ 
and carry the superconducting phase coherent dissipative current as in the case of a SC/I/SC junction\cite{baronetxt}. 
Moreover, $I_{\rm c2}$ in a long SC/NM/SC junction ($d>>\xi _{0}$) oscillates as a function of $d$ and $V$ with the period $ (4\pi E/v_{\rm F})d $ 
because of $\cos \left(\frac{2E}{v_{\rm F} }d \right)$ in eq. (\ref{ij2}). 
This behavior represents the phase shift by $\pi$ and the $0$-$\pi $ transition occurs as a function of $d$ and $V$. 
In the $V$-dependence of $I_{\rm c1}$, the divergence of $I_{\rm c1}$ at the gap voltage is also shown in the SC/I/SC junctions. 
Since, the density of states has a square-root singularity at the gap edge as we can see in eq. (\ref{ij1}), the amplitude of the Josephson critical current produces a logarithmic singularity at the gap voltage. 
This mechanism on the singularity is the same as that of SC/I/SC junction. 
In $I_{\rm c2}$, the current vanishes for $eV/2\Delta _{0}<1$ and jumps at the $eV/2\Delta _{0}=1$. 
This behavior is similar to that of the SC/I/SC junctions. 
In a long SC/NM/SC junction ($d>>\xi _{0}$), $I_{\rm c2}$ oscillates as a function of $V$ with the period $ (4\pi E/v_{\rm F})d $ 
as shown in eq. (\ref{ij2}). 

\begin{figure}[t]
\begin{center}
\includegraphics[width=6.5 cm]{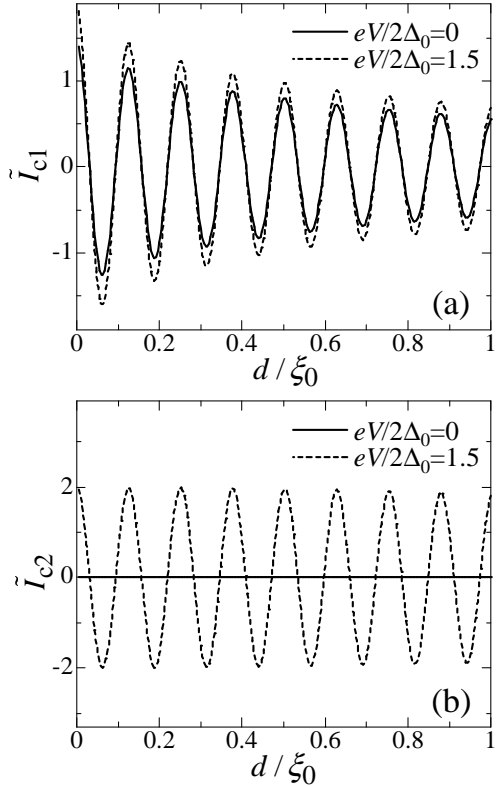}
\caption{(a) Normalized Josephson critical current, $\tilde I_{\rm c1}$, and (b) quasiparticle-pair-interference current, $\tilde I_{\rm c2}$, 
as functions of $d/\xi _{0}$ for $h_{\rm ex}/\Delta _{0}=50$ in SC/FM/SC junction, where $\tilde I_{\rm c1}=I_{\rm c1}\pi e/\sigma_{0} \Delta_{0}$ and 
$\tilde I_{\rm c2}=I_{\rm c2}\pi e/\sigma_{0} \Delta_{0}$. 
}
\label{icd_sfs}
\end{center}
\end{figure} 

\begin{figure}[t]
\begin{center}
\includegraphics[width=6.5cm]{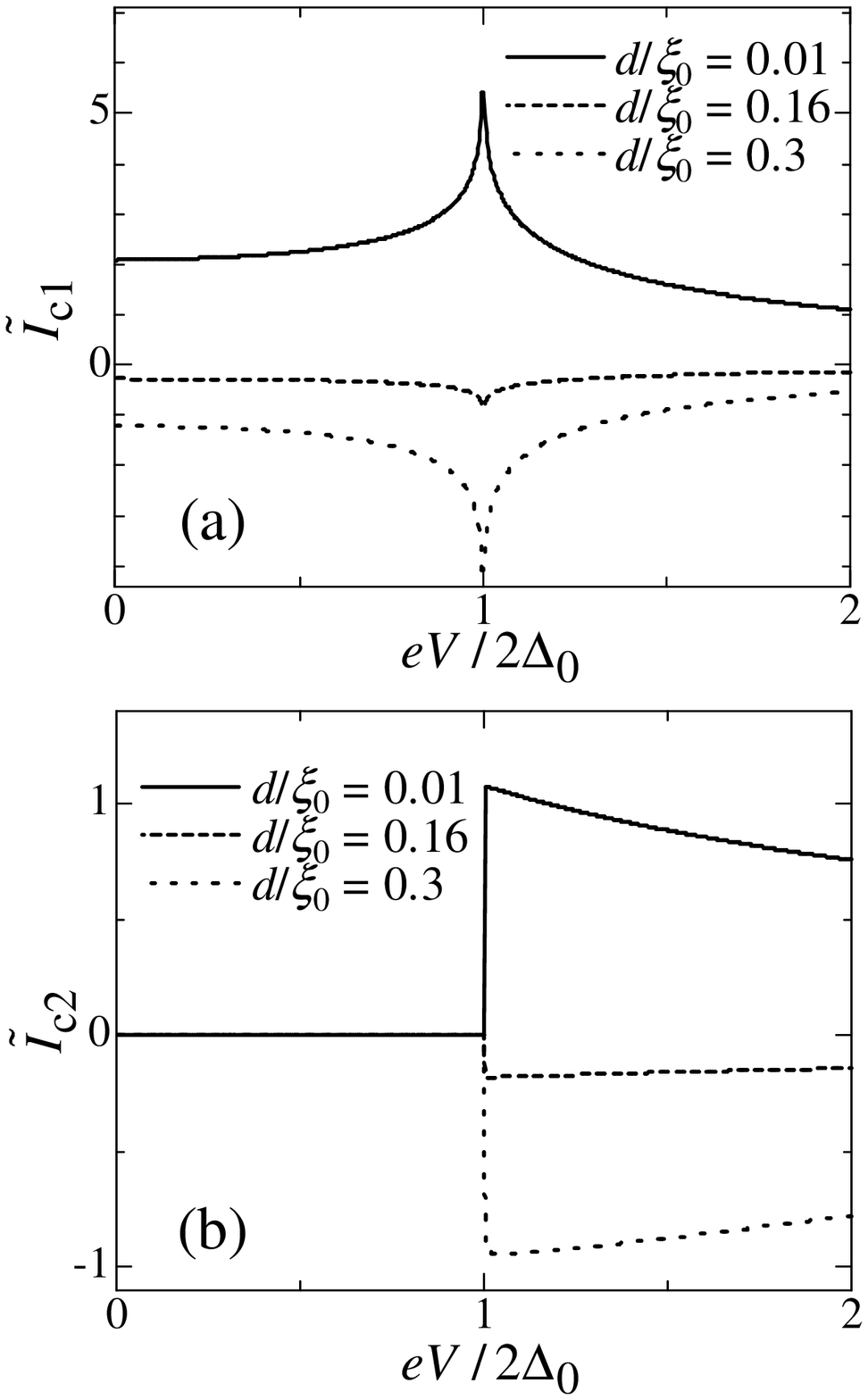}
\caption{(a) Normalized Josephson critical current, $\tilde I_{\rm c1}$, and (b) quasiparticle-pair-interference current, $\tilde I_{\rm c2}$, 
as functions of $eV/2\triangle _{0}$ for $h_{\rm ex}/\Delta _{0}=50$ in SC/FM/SC junction, where $\tilde I_{\rm c1}=I_{\rm c1}\pi e/\sigma_{0} \Delta_{0}$ and 
$\tilde I_{\rm c2}=I_{\rm c2}\pi e/\sigma_{0} \Delta_{0}$. 
}
\label{icv_sfs}
\end{center}
\end{figure}
\subsection{{\rm SC/FM/SC} junction}
In this subsection, the case of the SC/FM/SC junction is compared with that of the SC/NM/SC junction 
from the viewpoints of the $d$-dependences of $I_{\rm c1}$ and $I_{\rm c2}$ as shown in Fig.~\ref{icd_sfs}. In Fig.~\ref{icd_sfs} (a), the normalized $I_{\rm c1}$ is plotted as a function of $d/\xi _{0}$. 
For $eV/2\Delta _{0}=0$, i.e., dc Josephson effect, $I_{\rm c1}$ shows a damped oscillatory behavior as a function of $d$ 
and the 0-$\pi $ transition occurs by increasing $d$ unlike the case of SC/NM/SC junction. 
For $eV/2\Delta _{0}=1.5$, $I_{\rm c1}$ is qualitatively the same as that in the dc Josephson effect. 
On the other hand, $I_{\rm c2}$ is zero for $eV/2\Delta _{0}=0$, but  
 has a finite value for $eV/2\Delta _{0}=1.5$ similar to the SC/NM/SC junction. 
However, 
the $I_{\rm c2}$ changes its sign with increasing $d$ due to the 0-$\pi $ transition and vanishes at the transition point as shown in Fig.~\ref{icd_sfs} (b). 

Figure${~}$\ref{icv_sfs} shows the ac current amplitude as a function of $V$ in the SC/FM/SC junction. 
In Fig.${~}$\ref{icv_sfs} (a), the vertical axis is the normalized $I_{\rm c1}$ and the horizontal axis is the normalized voltage, $eV/2\Delta _{0}$. 
In this system, the Riedel peak exhibits a strong dependence on $d$ and changes its sign with increasing $d$ due to the 0-$\pi $ transition. 
Therefore, near the thickness at which the 0-$\pi $ transition occurs, the Riedel peak disappears as shown in Fig.~\ref{icv_sfs}(a). 
In Fig.${~}$\ref{icv_sfs} (b), it is found that $I_{\rm c2}$ has a finite value above $eV/2\Delta _{0}=1$. 
For $d/\xi _{0}=0.01$, the behavior of $I_{\rm c2}$ is the same as those of SC/I/SC and SC/NM/SC junctions. 
On the other hand, $I_{\rm c2}$ changes its sign with increasing $d$ due to the 0-$\pi $ transition and 
disappears near the thickness at which the 0-$\pi $ transition occurs, as shown in Fig.~\ref{icv_sfs} (b). 

The $I_{\rm c1}$ shows the damped oscillatory behavior as a function of $d$ and the 0-$\pi $ transition occurs. 
In eq. (\ref{ij1}), only the ratio of $h_{ex}$ and $v_{\rm F}$ determines the period of oscillation in the $I_{\rm c1}$-$d$ curve. 
The mechanism of 0-$\pi $ transition in the ac Josephson effect is the same as that of the 0-$\pi $ transition in the dc Josephson effect. 
Concerning $I_{\rm c2}$, it has a finite value above $eV/2\Delta _{0}=1$, similar to the SC/I/SC and SC/NM/SC junctions. 
The behavior of $I_{\rm c2}$ is quite different from that of SC/NM/SC junction. 
In the SC/FM/SC junction, the oscillating term in eq. (\ref{ij2}) is composed of two part. 
One is $\cos \left(2Ed/v_{\rm F}\right)$ (the region of $E$ being from 0 to $eV/2-\Delta_{0}$), the other is $\cos \left(2h_{ex}d/v_{\rm F}\right)$. 
In the practical case, $h_{ex}$ is experimentally larger than $eV$\cite{ryazanov_sfs1, kontos_sfs1, sellier_sfs1, sellier_sfs2, bauer_sfs1, frolov_sfs1, bell_sfs1, born_sfs1, weides_sfs1, 
oboznov_sfs1, shelukhin_sfs1}. 
Therefore, the period is dominated by $h_{ex}$ and is short compared with $\xi _{0}$. 
The $V$-dependence of $I_{\rm c1}$ and $I_{\rm c2}$ in the SC/FM/SC junction shows remarkable phenomenon. 
In $I_{\rm c1}$, the Riedel peak exhibits a strong dependence on $d$. 
It changes its sign with changing $d$ and disappears at the 0-$\pi $ transition point as shown in Fig. 6(a). 
$I_{\rm c2}$ also changes its sign with changing $d$ and disappears at the 0-$\pi $ transition point as shown in Fig. 6(b). 
We expect that these results provides a new method to observe the 0-$\pi $ transition in SC/FM/SC junctions.

\begin{figure}[t]
\begin{center}
\includegraphics[width=6.5cm]{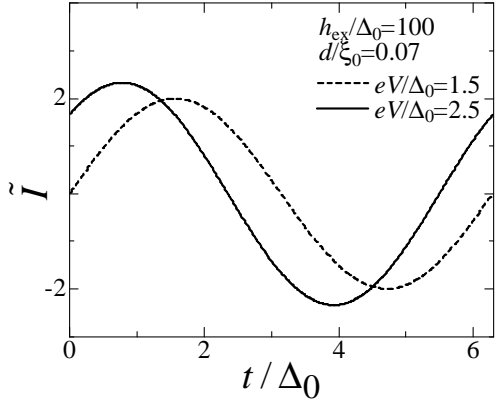}
\caption{ Ac current $\tilde I$ as a function of $t/\Delta _{0}$ for $h_{\rm ex}/\Delta _{0}=100$, 
$d/\xi _{0}=0.07$, $eV/\Delta _{0}=1.5$ and $2.5$. 
$\tilde I=I \pi e/\sigma_{0} \Delta_{0}$
}
\label{cpr}
\end{center}
\end{figure}

We take a look at the total ac current given by,
\begin{align}
I &=
		\sqrt{I_{\rm c1}^{2}+I_{\rm c2}^{2}}\sin\left(2eVt+\chi \right) \\
\chi &=\arctan\left(I_{\rm c2}/I_{\rm c1} \right)
\end{align}
Fig.~(\ref{cpr}) shows the current-phase relation(CPR) for $h_{\rm ex}/\Delta _{0}=100, d/\xi _{0}=0.07$, and $eV/\Delta _{0}=1.5$ and $2.5$. 
For $eV/2\Delta _{0}<1$, only the Josephson current flows because of $I_{\rm c2}=0$. 
Therefore, the CPR represents a conventional behavior as in a SC/I/SC junction as shown by the dashed line in Fig.~(\ref{cpr}). 
On the other hand, for $eV/2\Delta _{0}>1$,the phase of CPR is shifted by $\chi $ due to the finite QPIC. 
The behavior is shown in the solid line in Fig.~(\ref{cpr}). 
From these results, we can confirm the existence of the QPIC when the voltage dependence of the current-phase relation is measured. 

\section{Summary}
In summary, we have studied the ac Josephson effect in the SC/X/SC junction (X = NM, FM). 
Using a tunneling Hamiltonian and Green's function method, we obtained the Josephson current and 
the quasiparticle-pair-interference current (QPIC). 
The Josephson critical current, $I_{\rm c1}$, shows the Riedel peak at the gap voltage, $V=2\Delta /e$. 
In the SC/NM/SC junction, the Riedel peak exhibits weak dependence on $d$. 
The amplitude of QPIC, $I_{\rm c2}$, has a finite value above the gap voltage at $T=0 [\rm K]$. 
These behaviors are similar to those of SC/I/SC junctions. 
On the other hand, 
the critical currents in the SC/FM/SC junction show quite different behaviors compared with those in the SC/I/SC and SC/NM/SC junctions. 
In $I_{\rm c1}$, the Riedel peak exhibits a strong dependence on $d$ and changes its sign with increasing $d$ due to the 0-$\pi $ transition. 
Therefore, at the thickness of the 0-$\pi $ transition, the Riedel peak disappears. 
$I_{\rm c2}$ also shows the 0-$\pi $ transition with increasing $d$ and 
vanishes at the thickness of the 0-$\pi $ transition. 
The ac Josephson current has a strong dependence on the applied voltage, and 
at the gap voltage the amplitude of $I_{\rm c1}$ shows logarithmic divergence. 
The amplitude of the higher harmonic Josephson current also show the 
applied voltage dependence, similar to the Josephson current. 
The study of the ac Josephson effect in the SC/FM/SC junction gives 
a possibility
to observe the higher harmonic Josephson current. 

We have studied the ferromagnetic Josephson junction in the clean limit in this paper. 
In many experimental situations, FMs are usually in a diffusive transport region. 
However, the essence of 0-$\pi$ transition in the ac Josephson current may be the same even if the case of a dirty FM. 
Recently, the current-voltage characteristic was studied by using the superconducting phase difference 
coupled with the spin wave excitation in the ferromagnetic Josephson junction 
within the phenomenological model\cite{hikinoJPSJ}. 
And the dc Josephson current coupled with the spin wave excitation was also discussed\cite{houzet}. 
Therefore, it might be important to consider the spin wave excitation for the ac Josephson current from microscopic view point. 
These problems will be left in a future issue. 

\begin{acknowledgments}
This work is supported by a Grant in Aid for Scientific Research from MEXT, 
the next Generation Supercomputer Project of MEXT. 
The authors thank the Supercomputer Center, Institute for Solid State
Physics, University of Tokyo for the use of the facilities.
\end{acknowledgments}

\appendix
\section{Perturbative calculation of $I_{\rm c1}$ and $I_{\rm c2}$ in path integral formulation}
\subsection{Basic formula}
In the path integral formulation, the partition function is given by 
\begin{eqnarray}
Z = {\rm Tr[e}^{{\rm  - }\beta H} ] 
 	= \int {{\cal D}\Psi ^* {\cal D}\Psi {\rm e}^{- S} }, \nonumber
\end{eqnarray}
\begin{eqnarray}
    =\int {{\cal D}\Psi ^* {\cal D}\Psi \exp \left[ { - \int_0^\beta  {d\tau \left( {\int {dr\psi ^* \partial _\tau  \psi }  + H\left( \tau  \right)} \right)} } \right]} , 
\label{partition function} \nonumber \\
\end{eqnarray}
where Wick's rotation is performed, $it \rightarrow {\tau }, eV/2 \rightarrow  i{\rm \Omega }$ ($\beta = T^{-1}$, $\tau $ is an imaginary time variable, 
${\rm \Omega }$ is a temporary boson's Matsubara 
frequency. That becomes $eV/2$ by an analytic continuation.)  
and $\psi^* \partial _\tau  \psi \equiv \psi _{\rm L}^* \partial _\tau  \psi_{\rm L} +
\psi _{\rm FM}^* \partial _\tau  \psi_{\rm FM}+ \psi _{\rm R}^* \partial _\tau  \psi_{\rm R} $. It is noted that the energy of the left-hand superconductor (L)
is different from that of the right-hand superconductor (R), since the voltage, V, is applied. When these systems are separated each other, the time-development 
in the left-hand side is given by,
\begin{eqnarray}
U_{\rm L}\left(t,t' \right) &=&
	{\rm exp}\left[- i\left( {H_{\rm{L}}  - N_{\rm{L}} } \right)\left( {t - t'} \right) \right. \nonumber \\
        &+& \left. i\left( eV/2 \right)N_{\rm{L}} \left( {t - t'} \right)
\right],
\end{eqnarray}
while that in the right-hand side is 
\begin{eqnarray}
U_{\rm R}\left(t,t' \right) &=&
	{\rm exp}\left[ - i\left( {H_{\rm{R}}  - N_{\rm{R}} } \right)\left( {t - t'} \right) \right. \nonumber \\
        &-& \left. i\left( eV/2 \right)N_{\rm{R}} \left( {t - t'} \right) \right],
\label{unitary op}
\end{eqnarray}
Here, we make a transformation as, 
\begin{eqnarray}
 &&\psi _{\rm{L}} \left( {\textnormal{\mathversion{bold}$r$},t} \right) = 
		{\rm{e}}^{i\left( {e{\rm{V/2}}} \right)t} \tilde \psi _{\rm{L}} \left( {\textnormal{\mathversion{bold}$r$},t} \right),
\label{psi_l} \\ 
 &&\psi _{\rm{R}} \left( {\textnormal{\mathversion{bold}$r$},t} \right) = 
		{\rm{e}}^{ - i\left( {e{\rm{V/2}}} \right)t} \tilde \psi _{\rm{R}} \left( {\textnormal{\mathversion{bold}$r$},t} \right).
\label{psi_r}
\end{eqnarray}
Therefore, $H$ in the action, $S$, is transformed as
\begin{eqnarray}
\tilde H_{{\rm{BCS}}}^{\rm{L}}  &=&
\sum\limits_\sigma  {\int {dx
		\tilde \psi _{{\rm{L,}}\sigma }^* 
		\left[ {\partial _\tau   - \frac{1}{{2m}}\bigtriangledown^{2}  - \mu } 
		\right]
		\tilde \psi _{{\rm{L,}}\sigma } } }	\nonumber\\
		&& - g_{\rm{L}} \int {dx
		\tilde \psi _{{\rm{L,}} \uparrow }^* 
		\tilde \psi _{{\rm{L,}} \downarrow }^* 
		\tilde \psi _{{\rm{L,}} \downarrow } 
		\tilde \psi _{{\rm{L,}} \uparrow } },
\label{tr_bcs_hl} \\
\tilde H_{{\rm{FM}}} & =&
		\left(\rm{unchanged} \right),
\label{tr_fm_h} \\
\tilde H_{{\rm{BCS}}}^{\rm{R}} & =&
		\left( {{\rm{L}} \to {\rm{R}}} \right),
\label{tr_bcs_hr} \\
\tilde H_{\rm{T}}  &=& 
		\sum\limits_\sigma  {\int\limits_{\textnormal{\mathversion{bold}$r$} \in {\rm{L}},\textnormal{\mathversion{bold}$r$}' \in {\rm{FM}}} 
		{dxdx'{T_{x,x'}} {\rm{e}}^{ - ieVt/2} 
		\tilde \psi _{{\rm{L,}}\sigma }^* 
		\tilde \psi _{{\rm{FM,}}\sigma } } } \nonumber \\
		&+& \sum\limits_\sigma  {\int\limits_{\textnormal{\mathversion{bold}$r$} \in {\rm{R}},\textnormal{\mathversion{bold}$r$}' \in {\rm{FM}}} 
		{dxdx'{T_{x,x'}} {\rm{e}}^{ieVt/2} 
		\tilde \psi _{{\rm{R,}}\sigma }^* 
		\tilde \psi _{{\rm{FM,}}\sigma } } } \nonumber \\
&+& h.c.
\label{tr_th}
\end{eqnarray}
For convenience, the tilde is abbreviate. 
Where $\tilde \psi _{i, \sigma}^{(*)} \equiv \tilde \psi_{i, \sigma}^{(*)} \left(x \right)$, 
$i$ is L, R, or FM, 
$x = \left(\textnormal{\mathversion{bold}$r$},\tau \right)$, and ${T_{x,x'}} \equiv 
{T_{\bm r, \bm r'}} \delta (\tau -\tau ')$. 
The current operator is defined as,
\begin{eqnarray}
\hat J &=& 
		-e\frac{{dN_{\rm{L}} }}{{dt }} =  - e\int d {\bm r} \frac{dn_{\rm L}}{dt } 
		-ie\left[ {\int {dr} n_{\rm{L}} ,H} \right], \nonumber \\
		&=&
		- ie\sum\limits_\sigma  \int\limits_{\textnormal{\mathversion{bold}$r$} \in {\rm{L}},\textnormal{\mathversion{bold}$r$}' \in {\rm{FM}}} 
		d \textnormal{\mathversion{bold}$r$}d \textnormal{\mathversion{bold}$r$}'
		{T_{\bm r, \bm r'}} {\rm{e}}^{ - ieVt/2} \psi _{{\rm{L,}}\sigma }^\dag \psi _{{\rm{FM,}}\sigma } \nonumber \\ 
		&& + h.c., 
\label{tr_current op} \\
		n_{\rm{L}} &=&
		\sum\limits_\sigma  {\psi _{{\rm{L,}}\sigma }^\dag \psi _{{\rm{L,}}\sigma }}. 
\label{density of electron}
\end{eqnarray}

We eliminate the quartic interaction term in Hamiltonian eqs. (\ref{tr_bcs_hl}) and (\ref{tr_bcs_hr}) using Stratonovich-Hubbard transformation, 
\begin{eqnarray}
1 &=& \int {D\Delta ^* D\Delta \,\exp \left[ { - \frac{1}{g}\int {dx\left( {\Delta - g\psi _ \downarrow 
\psi _ \uparrow } \right)} \left( {\Delta ^* - g\psi _ \uparrow ^* \psi _ \downarrow ^* } \right)} \right]}, 
\label{sh_tr} \nonumber \\ 
\end{eqnarray}
where $\Delta^{(*)} \equiv\Delta^{(*)}(x) $. 
With this transformation, the action becomes,
\begin{eqnarray}
S &=&
S_{{\rm{cond}}}  + S_{{\rm{el}}} ,
\label{t_eff action} \\
S_{{\rm{cond}}}  &=& 
\int {dx} \frac{{\left| {\Delta _{\rm{L}}} \right|^2 }}{{g_{\rm{L}} }} + \int {dx} \frac{{\left| {\Delta _{\rm{R}}} \right|^2 }}{{g_{\rm{R}} }},
\label{cond_eff action} \\
S_{{\rm{el}}}  &=& 
\int {dx} \int {dx'} \Psi ^* \left( x \right)\left[ {G^ - } \right]\Psi \left( {x'} \right).
\label{eff action}
\end{eqnarray}
The Green's function, ${G^ - } \equiv G^{-}\left(x,x' \right)$, is a $6 \times 6$ matrix spanned both in the space of L, FM, R and Nambu space.
The electronic fields are expressed as, 
\begin{eqnarray}
&&\Psi \left(x \right)
	\equiv 
		\left( 
		\begin{array}{l}
 			\psi _{{\rm L} \uparrow } \\
 			\psi _{{\rm L} \downarrow }^* \\
 			\psi _{{\rm FM} \uparrow } \\
 			\psi _{{\rm FM} \downarrow }^* \\
 			\psi _{{\rm R} \uparrow } \\
 			\psi _{{\rm R} \downarrow }^*
 		\end{array}
 		\right),
\label{elc field op}
\end{eqnarray}
The inverse of the Green's function for electron system, $G^{-}$, is given by,
\begin{eqnarray}
\hat G^ -  &=& 
    \left[ {\begin{array}{*{20}c}
   { - \hat G_{\rm{L}}^ - } & {{{\hat T}}} & 0  \\
   {{\hat T}^{\rm{*}}} & { - \hat G_{{\rm{FM}}}^ - } & {{\hat T}}  \\
   {\rm{0}} & {{\hat T}^{\rm{*}} } & { - \hat G_{\rm{R}}^ - }  \\
\end{array}} \right], \nonumber \\
\label{g_function}
&=& - \hat G_0^ {-}  + {\rm{\hat P}}, \\
 - \hat G_0^{ - 1} &=& 
    \left[ {\begin{array}{*{20}c}
   { - \hat G_{\rm{L}}^ - } & 0 & 0  \\
   0 & { - \hat G_{{\rm{FM}}}^ - } & 0  \\
   {\rm{0}} & 0 & { - \hat G_{\rm{R}}^ - }  \\
\end{array}} \right], \\
\label{non_p g_function}
{\rm{\hat P}} &=& 
   \left[ {\begin{array}{*{20}c}
   0 & {{\hat T}} & 0  \\
   {{\hat T}^{\rm{*}} } & 0 & {{\hat T}}  \\
   {\rm{0}} & {{\hat T}^{\rm{*}}} & 0  \\
\end{array}} \right], \\
\label{p_matrix}
 - \hat G_{\rm{L}}^ - &=& 
                    \left[ {\partial _\tau  {\hat \sigma _{0}} -\left( { \frac{1}{{2m}}\Delta  + \mu } \right) \hat \sigma _3  
                    - \Delta _{\rm{L}} {\rm{e}}^{i\theta _{{\rm{L}}} \hat \sigma _3 } \hat \sigma _1 } \right] \nonumber \\
					&& \times \delta \left( {x - x'} \right), \\
\label{sc g_function}
- \hat G_{\rm R}^ - &=& 
					\left( {\rm L}\rightarrow {\rm R} \right), \\
- \hat G_{{\rm{FM}}}^ - &=& 
                    \left[ {\partial _\tau  {\hat \sigma  _{0}} - \left( { \frac{1}{{2m}}\Delta + \mu + i\sigma h_{ex} } \right)\hat \sigma _3 } \right] \nonumber \\
					&& \times \delta \left( {x - x'} \right), \\
\label{fm g_function}
{\hat T} &=& 
         \left[ {\begin{array}{*{20}c}
        {{\rm{e}}^{ - i\Omega \tau } {T_{\bm r, \bm r'}}} & 0  \\
        {\rm{0}} & { - {\rm{e}}^{i\Omega \tau } {T}_{\bm r, \bm r'}^{\rm{*}} 
		}  \\
\end{array}} \right]\delta (\tau -\tau ').
\label{t_matrix}
\end{eqnarray}
$\hat \sigma _{1}, \hat \sigma _{3}$ are the Pauli matrices and $\hat \sigma _{0}$ is a unit matrix. $\theta _{\rm L \left(\rm R \right)}$ is the 
phase of superconducting order parameter, $\Delta _{\rm L \left(\rm R \right)}\rm e^{i \theta _{\rm L \left(\rm R \right)}}$. Below, we consider the 
case that the superconductor L and R are same.

An auxiliary field, $\phi $, that couples to the current operator, $\hat J$, is introduced in eq. (\ref{tr_current op}) as, 
\begin{eqnarray}
S_{{\rm{cond}}}  &=& 
        \int {dx} \frac{{\left| {\Delta _{\rm{L}} } \right|^2 }}{{g_{\rm{L}} }} + 
        \int {dx} \frac{{\left| {\Delta _{\rm{R}} } \right|^2 }}{{g_{\rm{R}} }} \nonumber \\
		&&+\int_0^\beta  {d\tau \phi J}, \\
\label{eff action_ex}
\phi J &=& 
       \phi \int {dxdx'\Psi ^* \left( x \right)\left[ {\hat J} \right]} \Psi \left( {x'} \right), 
\label{phi_j}
\end{eqnarray}
where $\phi\equiv (\tau)$ and $J\equiv J \left(x,x' \right)$. 
The current operator is denoted as,
\begin{eqnarray}
\hat J &=& 
	\left[ {\begin{array}{*{20}c}
	0 & { - ie\hat R} & 0  \\
	{ie\hat R^* } & 0 & 0  \\
	0 & 0 & 0  \\
	\end{array}} \right], \\
\label{current op matrix}
\hat R &=& 
	\left[ {\begin{array}{*{20}c}
	{{\rm{e}}^{ - i\Omega \tau } T_{\bm r, \bm r'}} & 0  \\
	{\rm{0}} & {{\rm{e}}^{i\Omega \tau } {T}_{\bm r, \bm r'}^{\rm{*}} }  \\
	\end{array}} \right]\delta (\tau -\tau ').
\label{r_matrix}
\end{eqnarray}
Tracing out the electron fields from eq. (\ref{phi_j}), we obtain the effective free energy, $F_{\rm eff}$, with Josephson junction as following, 
\begin{eqnarray}
\beta F_{{\rm{eff}}}  &=& 
- {\rm{Tr}}\left\{ {\ln \left[ {\hat G^{ - 1} + 
\phi \hat J} \right]} \right\}, \nonumber \\
 &=&  
- {\rm{Tr}}\left\{ {\ln \left[ {\left( { - \hat G_0^{-} + 
\phi \hat J} \right) + {\rm{\hat P}}} \right]} \right\}.
\label{eff free energy}
\end{eqnarray}
In eq. (\ref{eff free energy}), Tr means taking trace with respect to the matrix element of the Green's function and integrating with respect to $x, x'$.
In the fourth order perturbation theory about the tunneling matrix, the current is approximated to be, 
\begin{eqnarray}
\left\langle J \right\rangle  &=& 
		\left. {\frac{{\partial F_{{\rm{eff}}} }}{{\partial \phi }}} \right|_{\phi  = 0}  = 
		- \frac{1}{\beta }\left. {\frac{\partial }{{\partial \phi }}{\rm{Tr}}\left\{ {\ln \left[ {\left( { - \hat G_0^{-} 
		+ \phi \hat J} \right) + {\rm{\hat P}}} \right]} \right\}} \right|_{\phi  = 0}, \nonumber \\
& \simeq& 
		\frac{1}{\beta }{\rm{Tr}}\left[ {\hat J \hat G_0 } \right] + 
		\frac{1}{\beta }{\rm{Tr}}\left[ {\hat J \hat G_0 {\rm{\hat P}}
		\hat G_0 } \right] \nonumber \\
		 &&+ \frac{1}{\beta }{\rm{Tr}}\left[ {\hat J \hat G_0 {\rm{\hat P}}
		\hat G_0 {\rm{\hat P}} \hat G_0 {\rm{\hat P}}
		\hat G_0 } \right].
\label{expected j op}
\end{eqnarray}
The first and second terms doesn't contribute the Josephson current, whose leading term is the third one in eq. (\ref{expected j op} ). 

Because Tr is not affected by a change of basis, the matrix element can be transformed from ({\mathversion{bold}$ r $},$ \tau $) component 
to its Fourier component ({\mathversion{bold}$ k $},$i \omega _{n}$). Here, $i \omega _{n}$ is the Matsubara frequency of the electrons. 
Here we made use of the fact that the tunneling matrix element differs from 0 only for 
$\textnormal{\mathversion{bold}$r$}\sim \textnormal{\mathversion{bold}$r$}' $and in the neighborhood of the SC/FM boundary, 
i.e., ${T_{\bm r, \bm r'}}=
{T_{0}}\delta \left(\textnormal{\mathversion{bold}$r$}-\textnormal{\mathversion{bold}$r$}'\right)
\delta \left(\textnormal{\mathversion{bold}$r$}-\textnormal{\mathversion{bold}$r$}_{\rm L (R)}\right)$, 
where $\textnormal{\mathversion{bold}$r$}_{\rm L (R)}$ 
is the position of the interface between the left (right) SC and the FM. 
Tracing out the matrix element of the Green's function, we can obtain the current formula as following, 
\begin{eqnarray}
I_{\rm J} &=&
		- 8eT^{\rm{4}}_{0} {\rm{Im}}\frac{1}{\beta }\sum\limits_{\left\{\textnormal{\mathversion{bold}$k$} \right\},i\omega _n } 
		{\left[ {{\rm{e}}^{i\left( {\textnormal{\mathversion{bold}$k$}_{{\rm{FM}}}^ \uparrow   
		- \textnormal{\mathversion{bold}$k$}_{{\rm{FM}}}^ \downarrow  } \right) \textnormal{\mathversion{bold}$r$}} } \right.}  \nonumber \\ 
		&\times & f_{\rm{L}} \left( {\textnormal{\mathversion{bold}$k$}_{\rm{L}} ,i\omega _n  + i\Omega } \right)
		g_{{\rm{FM,}} \uparrow }^{\left( {\rm{0}} \right)} 
		\left( {\textnormal{\mathversion{bold}$k$}_{{\rm{FM}}}^ \uparrow  ,i\omega _n } \right) \nonumber \\ 
		&\times & \left. { g_{{\rm{FM,}} \downarrow }^{\left( {\rm{0}} \right)} 
		\left( {\textnormal{\mathversion{bold}$k$}_{{\rm{FM}}}^ \downarrow  , - i\omega _n } 
		\right)f_{\rm{R}}^* \left( {\textnormal{\mathversion{bold}$k$}_{\rm{R}} ,i\omega _n  - i\Omega } \right)} \right],
\label{t_ij1}
\end{eqnarray}
where $\textnormal{\mathversion{bold}$r$}=\textnormal{\mathversion{bold}$r$}_{\rm R}-\textnormal{\mathversion{bold}$r$}_{\rm L}$ and 
each Green's function is given by 
\begin{eqnarray}
&&f_{{\rm L}} (\textnormal{\mathversion{bold}$k$}_{\rm L} ,i\omega_{n}+i\Omega  ) 
= \frac{\Delta _{\rm L}\rme^{i \varphi_{\rm L}}}
         {\left(\omega_{n}+\Omega \right)^{\rm 2}  + \xi_{\rm L}^2  + \Delta_{\rm L}^2},~~~~~\\
&&f_{{\rm R}} (\textnormal{\mathversion{bold}$k$}_{\rm L} ,i\omega_{n}-i\Omega  )
=\left( {\rm L}\rightarrow {\rm R} \right), \\        
&&g_{{\rm FM},\uparrow}^{(0)}  (\textnormal{\mathversion{bold}$k$}_{\rm FM \uparrow } ,i\omega_n ) 
=
  - \frac{1}{{i\omega_n  - \xi _{{\rm FM}}^{\uparrow}  }},\\
&&g_{{\rm FM},\downarrow}^{(0)}  (-\textnormal{\mathversion{bold}$k$}_{\rm FM \downarrow}  ,-i\omega_n ) 
=
  - \frac{1}{{i\omega_n  + \xi _{{\rm FM}}^{\downarrow}}},\\
&&\xi _{{\rm L(R)}}
 = \frac{{\klr^{\rm 2} }}{2m} - \mu,  \\ 
&&\xi _{{\rm FM}}^\sigma
= \frac{{\km^2}}{2m} - \mu  - \sigma h_{{\rm ex}}.
\end{eqnarray}
Note that $k_{\rm FM \uparrow }$ and $k_{\rm FM \downarrow}$ are independent of each other. 
After integrating Green's functions of the FM as to $k_{\rm FM,\sigma }$, we can obtain the following formula, 
\begin{eqnarray}
I_{\rm J} &=&
		16eT_{0}^{4} \left[ \frac{mv}{2 \pi d} \right]^{2}
		{\rm Im} \left[ i \frac{\Delta ^{2}}{\beta} \sum_{k_{\rm L}, k_{\rm R}, \omega_{n}>0} \right. \nonumber \\
		&\times & \frac{1}{(\omega_{n}+\Omega)^{2}+E_{k_{\rm L}}^{2}}
		\frac{1}{(\omega_{n}-\Omega)^{2}+E_{k_{\rm R}}^{2}} \nonumber \\
		&\times & \left.\cos \left(\frac{2h_{ex}}{v_{\rm F}d} \right) 
		e^{-2 \omega_{n}d/v_{\rm F}} \right] \sin\theta \nonumber \\
		&+ &
		16eT_{0}^{4} \left[ \frac{mv}{2 \pi d} \right]^{2}
		{\rm Im} \left[ \frac{\Delta ^{2}}{\beta} \sum_{k_{\rm L}, k_{\rm R}, \omega_{n}>0} \right. \nonumber \\
		&\times & \frac{1}{(\omega_{n}+\Omega)^{2}+E_{k_{\rm L}}^{2}}
		\frac{1}{(\omega_{n}-\Omega)^{2}+E_{k_{\rm R}}^{2}} \nonumber \\
		&\times & \left. \cos \left(\frac{2h_{ex}}{v_{\rm F}d} \right) 
		e^{-2 \omega_{n}d/v_{\rm F}} \right] \cos\theta, \nonumber \\
		&\equiv & I_{\rm c1} \sin \theta + I_{\rm c2} \cos \theta, 
\label{t_ij2}
\end{eqnarray}
where $v$ is a volume of the FM, $k_{\rm F}$ is the Fermi wave number, $\beta =T^{-1}$.

\begin{figure}[t]
\begin{center}
\includegraphics[width=7.5cm]{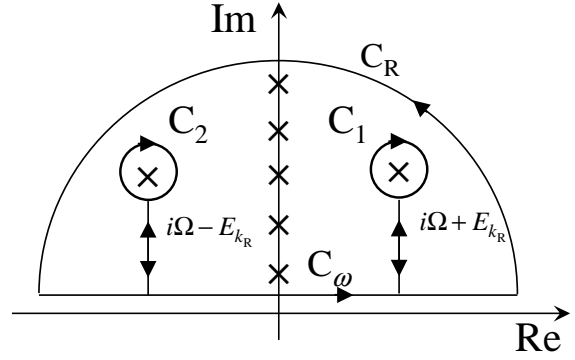}
\caption{Paths in the complex plane for the contour integration of eq. (B1)}
\label{figb}
\end{center}
\end{figure}
 
\section{Summation of Matsubara frequency}
We evaluate Matsubara frequency summation in (\ref{t_ij2}). This summation is performed by a contour integration in Fig. B. 
The integral has the form 
\begin{eqnarray}
I_{\rm C} &=&
		-\oint_{\rm{C}} {\frac{dz}{2\pi i}f(z)n_{\rm F}(z)e^{i 2zd/v_{\rm F}}}, \\
f(z) 
				  &=&
                       \frac{1}{{z + i\Omega  - E_{k_{\rm{L}} } }}
				       \frac{1}{{z + i\Omega  + E_{k_{\rm{L}} } }} \nonumber \\
				       &\times &
				       \frac{1}{{z - i\Omega  - E_{k_{\rm{R}} } }}
				       \frac{1}{{z - i\Omega  + E_{k_{\rm{R}} } }}, \\
{\rm C}
				&=&
				       {\rm C}_{1}+{\rm C}_{2}+{\rm C}_{\omega }+{\rm C}_{\rm R},
\end{eqnarray}
where $n_{\rm F}(z)$ is a Fermi distribution function,  ${\rm C}_{1}$ and ${\rm C}_{2}$ are a closed path enclosing the pole of the anomalous Green's function, 
${\rm C}_{\omega }$ is a integration on the real axis of complex plane, and ${\rm C}_{\rm R}$ is a large semicircle of radius $R$ in the limit as 
$R\rightarrow \infty $. The contour integration with ${\rm C}_{\rm R}$ is zero because of the Jordan's theorem. When each contour integration is performed 
, the following result is obtained, 
\begin{eqnarray}
I_{\rm C1} &=&
		\frac{1}{4E_{k_{\rm L}}E_{k_{\rm R}}} \nonumber \\
		&\times &
		\left(\frac{1}{i2\Omega +E_{k_{\rm R}}-E_{k_{\rm L}}}-
		\frac{1}{i2\Omega +E_{k_{\rm R}}-E_{k_{\rm L}}} \right) \nonumber \\
		&\times & n_{\rm F}\left(E_{k_{\rm R}}\right) \exp\left( i\frac{i2\Omega +2E_{k_{\rm R}}}{v_{\rm F} }d \right), \\
I_{\rm C2} &=&
		-\frac{1}{4E_{k_{\rm L}}E_{k_{\rm R}}} \nonumber \\
       &\times &
		\left(\frac{1}{i2\Omega -E_{k_{\rm R}}-E_{k_{\rm L}}}-
		\frac{1}{i2\Omega -E_{k_{\rm R}}+E_{k_{\rm L}}} \right) \nonumber \\
		&\times &n_{\rm F}\left(-E_{k_{\rm R}}\right) \exp\left( i\frac{i2\Omega -2E_{k_{\rm R}}}{v_{\rm F} }d \right), \\
I_{\omega } &=&
\left[ 		
		-\int_{ - \infty }^\infty  {\frac{d\omega }{\pi i}}
		\frac{1}{4E_{k_{\rm L}}E_{k_{\rm R}}}
		\frac{1}{i2\Omega -E_{k_{\rm R}}-E_{k_{\rm L}}} \right. \nonumber \\
		&\times &\left(\frac{1}{i2\Omega +2\omega -2E_{k_{\rm L}}}+\frac{1}{i2\Omega -2\omega -2E_{k_{\rm R}}} \right)  \nonumber \\
		&+& \int_{ - \infty }^\infty  {\frac{d\omega }{\pi i}}
		\frac{1}{4E_{k_{\rm L}}E_{k_{\rm R}}}
		\frac{1}{i2\Omega +E_{k_{\rm R}}-E_{k_{\rm L}}} \nonumber \\
		&\times & \left(\frac{1}{i2\Omega +2\omega -2E_{k_{\rm L}}}+\frac{1}{i2\Omega -2\omega +2E_{k_{\rm R}}} \right) \nonumber \\
		&+& \int_{ - \infty }^\infty  {\frac{d\omega }{\pi i}}
		\frac{1}{4E_{k_{\rm L}}E_{k_{\rm R}}}
		\frac{1}{i2\Omega -E_{k_{\rm R}}+E_{k_{\rm L}}} \nonumber \\
		&\times & \left(\frac{1}{i2\Omega +2\omega +2E_{k_{\rm L}}}+\frac{1}{i2\Omega -2\omega -2E_{k_{\rm R}}} \right) \nonumber \\
		&-& \int_{ - \infty }^\infty  {\frac{d\omega }{\pi i}}
			\frac{1}{4E_{k_{\rm L}}E_{k_{\rm R}}}
			\frac{1}{i2\Omega +E_{k_{\rm R}}+E_{k_{\rm L}}} \nonumber \\
			&\times &\left. \left(\frac{1}{i2\Omega +2\omega +2E_{k_{\rm L}}}+\frac{1}{i2\Omega -2\omega +2E_{k_{\rm R}}} \right) 
		 \right] \nonumber \\
		&&\times n_{F}\left(\omega \right) \exp\left(i \frac{2 \omega }{v_{\rm F} }d \right).
\end{eqnarray}
Here, we consider the case of absolute zero temperature. Then $n_{\rm F}\left(E_{k_{\rm L(R)}} \right)=0, 
n_{\rm F}\left(\omega \right)=\Theta (-\omega )$. Making an analytic continuation $i2\Omega \rightarrow eV+i\delta $ and integrations with respect to 
$\omega $ and $E_{k_{\rm L}}$ or $E_{k_{\rm R}}$, we obtain the analytical form of $I_{\rm c1}$ and $I_{\rm c2}$,
\begin{align}
I_{\rm c1} &=
		\frac{\sigma _{0}\Delta_{0} ^{2}}{\pi e}
		\int_{\Delta_{0} }^\infty dE \nonumber \\
		&\times \frac{\Theta \left(\Delta_{0} -\left| E-eV\right| \right)}{\sqrt{E^{2}-\Delta_{0} ^{2}}\sqrt{\Delta_{0} ^{2}-\left(E-eV \right)^{2}}} \nonumber \\
		&\times \left\{ {\rm Ci}\left( \frac{2E-eV}{v_{\rm F}}d \right) \sin \left( \frac{2E-eV}{v_{\rm F}}d \right) \right. \nonumber \\
		&\left.-\cos \left(\frac{2E-eV}{v_{\rm F}}d \right) \left[{{\rm Si}\left(\frac{2E-eV}{v_{\rm F}}d \right)- \frac{\pi }{2}} \right] \right\} \nonumber \\
		&\times \cos \left(\frac{2h_{\rm ex}}{v_{\rm F}}d \right),\\
I_{\rm c2} &=
		\frac{\sigma_{0}\Delta_{0} ^{2}}{e}
		\int_{0}^{eV/2-\Delta_{0}} {dE} \nonumber \\
		&\times \frac{\Theta \left(eV-2\Delta_{0} \right)}
		{\sqrt{\left(E+eV/2 \right)^2-\Delta_{0} ^{2}}\sqrt{\left(E-eV/2 \right)^2-\Delta_{0} ^{2}}} \nonumber \\
		&\times \cos \left(\frac{2E}{v_{\rm F}}d \right) \cos \left(\frac{2h_{\rm ex}}{v_{\rm F} }d \right),
\end{align}
where $\sigma _0  = 16\pi e^{\rm{2}} {\rm{T}}_{\rm{0}}^{\rm{4}} N_{\rm{L}} (0)N_{\rm{R}} (0)\left[{ mv/(2\pi d)}
 \right]^{\rm{2}}$.


\end{document}